\documentclass[aps,prb,epsfig,showpacs,superscriptaddress,amsmath,amssymb,twocolumn]{revtex4-1}
\usepackage{multirow}
\usepackage{amssymb}
\usepackage{amsmath}
\usepackage{graphicx}
\usepackage{epsfig}
\usepackage{dcolumn}
\usepackage{bm}
\usepackage{ulem}
\usepackage{color}
\usepackage{soul}
\usepackage{braket}
\graphicspath{ {./figs/} }
\newcommand{\dw}{{\downarrow }}
\newcommand{\up}{{\uparrow }}

\newcommand{\be}{\begin{equation}}
\newcommand{\ee}{\end{equation}}
\newcommand{\ba}{\begin{eqnarray}}
\newcommand{\ea}{\end{eqnarray}}
\newcommand{\baa}{\begin{eqnarray*}}
\newcommand{\eaa}{\end{eqnarray*}}
\newcommand{\dg}{^{\dagger}}

\newcommand{\bk}{{\bf k}}
\newcommand{\bz}{{\bf z}}
\newcommand{\bdk}{\hat {d} ({\bk })}
\newcommand{\bnk}{\hat {n} ({\bk} )}

\newcommand{\bpm}{\begin{pmatrix}}
\newcommand{\epm}{\end{pmatrix}}

\newlength{\fight}
\fight=\columnwidth

\begin{document}

\title{$^{3}$He-R: A Topological $s^{\pm}$ Superfluid with Triplet Pairing}
\author{T.~Tzen Ong}
\affiliation{Center for Materials Theory, Department of Physics \& Astronomy, Rutgers University, Piscataway NJ 08854, USA}
\author{Piers Coleman}
\affiliation{Center for Materials Theory, Department of Physics \& Astronomy, Rutgers University, Piscataway NJ 08854, USA}
\affiliation{Department of Physics, Royal Holloway, University of London, Egham, Surrey TW20 0EX, UK}
\date{\today}

\begin{abstract}
We show that when spin and orbital angular momenta are entangled by spin-orbit coupling,
this transforms a topological spin-triplet superfluid/superconductor state, such as $^{3}$He-B,
into a topological $s^{\pm}$ state, with non-trivial gapless edge states. Similar to $^{3}$He-B, the $s^{\pm}$ state
also minimizes on-site Coulomb repulsion for weak to moderate interactions. A phase transition into
a topological $d$-wave state occurs for sufficiently strong spin-orbit coupling.
\end{abstract}
\maketitle

\section{Introduction}
Topological states of matter, including topological insulators,
superconductors and superfluids are of great
current interest\cite{Volovik2003,RyuLudwigPRB2008,RyuLudwigNJOP2010,QiZhangRMP2011}.
Spin-orbit coupling plays a key role in driving
the non-trivial topology of 3D topological insulators, and a
topological superconducting phase (spinless $p + ip$) can also be
induced by proximity effect between a conventional $s$-wave
superconductor and a material with strong spin-orbit coupling, such as
a topological insulator \cite{KaneLiangPRL2008,SarmaSauPRL2010,OregFelixPRL2010}.
In this paper, we will show that a topological $s^{\pm}$ state can also be
generated using the converse effect of spin-orbit coupling on a $p$-wave
condensate.

To illustrate this physics, we introduce a toy model, describing 2D
$^{3}$He-B with an additional tunable Rashba coupling. This tunable
coupling term is absent in real He-3, but the model provides a
simple and pedagogical example of the effect of strong spin-orbit coupling on a
topological superconductor that may be generalized to a larger class
of superconductors, such as Sr$_{2}$RuO$_{4}$ \cite{RiceSigrist1995,MaenoNature1998,
SigristNature1998, MacKenzieMaenoRMP2003}, in which either spin, or
some other internal degree of freedom may become entangled with the
momentum-space structure of the condensate. $^{3}$He-B is the canonical
example of a topological superfluid\cite{VollhardtWolfle2013}.  An early theory of p-wave
pairing applicable to the B-phase of He-3B, was proposed by Balian and
Werthammer in 1963\cite{BalianWerthamerPhysRev1963}, prior to its
experimental discovery in the 1970's
\cite{OsheroffRevModPhys1997,LeggettRMP1975,WheatleyRMP1975,VollhardtWolfle2013}.
While the anisotropic $p$-wave nature of its pairing due to the
fermionic hard-core repulsion was predicted early on
\cite{BalianWerthamerPhysRev1963,AndersonPhysRev1961}; the underlying
topological character of the wavefunction, together with its gapless
Majorana edge states were only pointed out in 2003 by Volovik
\cite{Volovik2003,VolovikJTEP2009a,VolovikJETP2009b}; more recent
works have connected He-3B with a much more general class of
topological superfluids\cite{QiZhangPRL2009, WuSaulArXiv:1308.4436}.

\begin{figure}
\begin{center}
\includegraphics[width=\columnwidth]{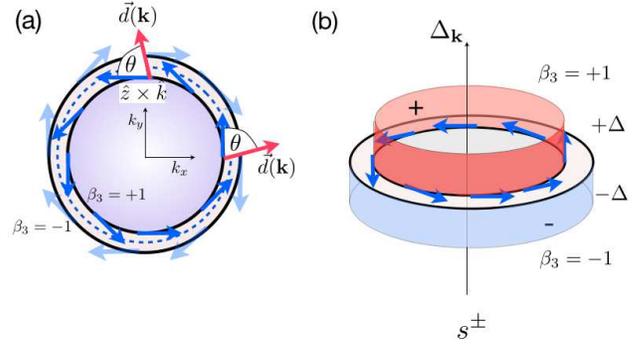}
\end{center}
\caption{(a) With Rashba coupling, the 2D Fermi surface is split into two with opposing
helicities $\beta_{3}= \pm 1 $. The relative orientation of the helicity vector
$\hat{z} \times \hat{k}$ and the triplet pairing $\hat d(\bk)$ vector
is $\theta $. (b) In the superfluid $^{3}$He-R condensate, the gap is
maximized  when the helicity and $\hat d(\bk)$ vectors align ($\theta
=0$), developing an $s^{\pm }$ gap function with opposite signs on the two Fermi
surfaces.}
\label{fig:2D SOC Bands}
\end{figure}

\begin{figure}
\begin{center}
\includegraphics[width=\columnwidth]{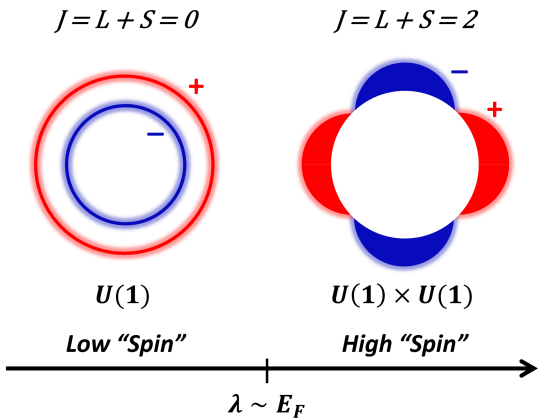}
\end{center}
\caption{A phase transition from the $s^{\pm}$ state into a $d$-wave state occurs when the spin-orbit interaction is sufficiently strong to lift one of the helical bands above $E_F$.}
\label{Fig:Phase transition}
\end{figure}

$^{3}$He-B is a $p$-wave superfluid with unbroken time-reversal
symmetry. Although the underlying gap functions contain nodes, the
combination of orthogonal spin channels ($\sigma_{x,y,z}$) causes
the various p-wave gaps to add in quadrature, hiding one-another's
nodes and giving rise to a fully gapped excitation spectrum.
In the absence of spin-orbit
coupling, the spin ($S$) and angular momentum ($L$) of the Cooper
pairs are well-defined quantum numbers. However, spin-orbit coupling
entangles $L$ and $S$, and only the total angular momentum, $J = L +
S$, is well-defined. We show that when orbital and spin angular momentum become mixed,
a $p$-wave superfluid is transformed into a topological
$s^{\pm}$ ($J = 1-1= 0$) or a nodal $d$-wave ($J =1+1= 2$) superfluid,
as the spin-orbit coupling strength is increased.

Our analysis includes the $U(1)$ rotational
degree of freedom between the spin-orbit, $\hat{n}_{\bk}$,
and superconducting $\hat{d}_{\bk}$ vectors, which was 
ignored in previous works \cite{FrigeriSigristPRL2004,SatoSatoshiPRB2009,TanakaSatoPRL2010}
on non-centro-symmetric superconductors, where it was
assumed that $\hat{n}_{\bk}$ and $\hat{d}_{\bk}$ are always parallel
due to strong spin-orbit coupling. Here, we show that the strong Coulomb repulsion
breaks the alignment of $\hat{n}_{\bk}$ and $\hat{d}_{\bk}$, and the
mixing of $s$ and $d$-wave spin-singlet pairing, with the 
$p$-wave spin-triplet pairing naturally arises from 
the in-phase and counter-phase rotation of
$\hat{n}_{\bk}$ and $\hat{d}_{\bk}$ respectively.

Specifically, our key results are: \begin{enumerate}
\item
At weak to moderate spin-orbit coupling, the ground-state of
isotropic $^{3}$He-B adiabatically transforms into a ``low-spin''
$J= 1-1 =0$ s-wave condensate, made up of two fully gapped
spin-polarized Fermi surfaces of opposite pairing phase. This state
retains the topological character of its p-wave parent, forming an
``$s^{\pm}$'' state with topologically protected gapless edge states.

\item In the presence of strong spin-orbit coupling ($\lambda_{\bk} \approx \mu$),
the system undergoes a topological phase transition into a ``high-spin''
topological $d$-wave state with angular momentum ($J =1+1= 2$). We note that the
$d$-wave state has been discussed in the context of neutron stars by
earlier groups \cite{PinesAlparNature1985,RudermanPRL1970}, although
the topological nature of the $d$-wave state was not appreciated.

\end{enumerate}

Our results show that an apparently s-wave
superfluid/superconductor can hide
pairing in a higher angular momentum channel, thereby
minimizing a hard-core repulsion or a local Hubbard repulsion.
\be \label{eqn:Hubbard constraint}
\lim_{U \rightarrow \infty} \langle
c_{\uparrow} (\vec{x}) c_{\downarrow} (\vec{x}) \rangle = 0
\ee

The breaking of inversion symmetry (${\mathcal{I}}$) mixes 
even-parity spin-singlet and odd-parity spin-triplet
Cooper pairs in non-centosymmetric superconductors,
and the effects of $s$-wave and $d$-wave pairing in the presence of
strong Coulomb repulsion $U > \lambda$, with a resulting 
``low" spin to ``high" spin phase transition is  addressed in Sec.~\ref{sec:Coulomb}.

While the strong spin-orbit coupling necessary for a ``low-spin'' to
``high-spin'' transition is un-physical in actual
$^{3}$He-B, it may be realized in cold-atom systems
\cite{LinSpielmanNature2011}. Another interesting possibility is the
iron-based systems which have strong orbital exchange hoppings, where
orbital iso-spin ($I$) plays a similar role to spin in $^{3}$He-B
\cite{OngColemanPRL2013}, allowing us to generate ($J = L + I = 0$)
$s^{\pm}$ or ($J = L + I = 4$) $g$-wave superconducting states.

\section{$^{3}$$\hbox{He}$-R: two dimensional  $^{3}$He-B with Spin-Orbit Coupling}
\label{sec:He3R}

We now formulate a simple model of two dimensional $^{3}$He-B with a
a Rashba spin-orbit coupling  that we  refer to as $^{3}$He-R.
A Rashba coupling is introduced into the kinetic energy,
by replacing $\epsilon_{\bk }\rightarrow \epsilon_{\bk }+ \lambda
(\hat {\bf z} \times  {\bf k})\cdot \vec{\sigma }$, where $\hat {\bf
z}$ is normal to the plane.  The Rashba term is absent in real
$^{3}$He-B, but might be realized in other contexts, such as a cold-atom system.
The toy model for $^{3}$He-R is then
\begin{eqnarray}
\label{eq:H0 SOC}
H  &=&  \sum_{\bk } c\dg_{\bk } \left[ \epsilon_{\bk }
+ \lambda_{\bk} \bnk \cdot \vec{\sigma }
\right] c_{\bk }\cr
 &+&
\sum_{\bk \in \frac{1}{2}\hbox{MS}}
\left[ \Delta c\dg_{\bk } (\bdk \cdot \vec{\sigma }) i
  \sigma_{2} c\dg_{-\bk } + {\rm H.c}\right],
\end{eqnarray}
where the summation for the pairing term is over half of momentum
space (MS), most simply implemented by restricting $k_{x}>0$.  Here
$\bnk = \hat \bz \times  \hat \bk $ denotes the direction of the
Rashba field, $c\dg_{\bk }\equiv (c\dg_{\bk \up},c\dg_{\bk \dw}) $
is the electron creation operator and $\hat d(\bk)$ is the d-vector
determining the local direction of p-wave pairing in momentum
space. Here we have restricted ourselves to the class of Balian-Werthammer
p-wave condensates in which the d-vector is of constant magnitude.
We shall follow the normal convention of choosing
$\lambda_{\bk} = \lambda |\vec{k}|$, but will adopt
a simpler, momentum-independent interaction, $\lambda_{\bk} = \lambda$ in
Sec.~\ref{sec:Coulomb} to illustrate the qualitative effects of
a hard-core/Coulomb repulsion.

Following Balian and Werthamer, we write the Hamiltonian in Nambu notation,
\be
\label{rash2}
H = \sum_{\bk \in \frac{1}{2}\hbox{MS}} \psi \dg_{\bk }{\cal H}_{\bk }\psi_{\bk }
\ee
\begin{eqnarray}
\label{rash3}
{\cal H}_{\bk } = (\epsilon_{\bk } + \lambda_{\bk } \bnk \cdot \vec{\sigma} )
\gamma_{3} + (\Delta \bdk \cdot \vec{\sigma} )\gamma_{1}.
\end{eqnarray}
Here $\vec{\gamma}= (\gamma_{1},\gamma_{2},\gamma_{3})$
denotes the three Nambu matrices and
\begin{equation}
\label{rash4}
\psi_{\bk} = \left(\begin{matrix} c_{\bk \uparrow}\cr c_{\bk
\downarrow}
\cr c\dg _{-\bk \downarrow }\cr -c\dg _{-\bk\uparrow }
\end{matrix} \right)
\end{equation}
is the Balian-Werthammer four-component spinor.
Two dimensional $^{3}$He-B is described by the case where $\lambda_{\bk }=0$.

In this case, the d-vector wraps around the Fermi surface, and can be
written in the general form $\hat d (\bk ) = O\cdot (\hat k_{x},\hat
k_{y})$ where $O$ is a two dimensional orthogonal matrix; the cases
$\det (O)= \pm 1$ correspond to a $\hat d$ vector
that winds in the same, or opposite sense to the Rashba vector $\bnk$.
Consider the case where
$\hat d (\bk )= \bnk $, so that
\begin{eqnarray}
\label{eq:d vector}
\hat d (\bk )\cdot \vec{\sigma } & = & -\hat k_{y} \sigma_{x}+ \hat k_x \sigma_y,
\end{eqnarray}
corresponding to a $d$-vector that points tangentially in momentum
space.
The corresponding paired state is fully gapped, with spectrum
\be
E_{\bk }= \sqrt{\epsilon_{\bk }^{2}+ \Delta^{2}}.
\ee

The B-phases of He-3
have topological character captured by the fact that the $\hat d (\bk )$ has a finite
winding number $n = \pm 1$ in spin space, where
\be
n = \oint \hat{z} \cdot \left( \bdk\dg  \times \partial_{a} \bdk \right) \frac{d k_{a}}{2\pi} = \pm 1.
\label{eq:winding number}
\ee
The fully gapped structure of the spectrum hides the underlying p-wave
nodes and the topological character.

We now re-introduce the spin-orbit coupling term
$\lambda_{\bk} (\bnk \cdot \vec\sigma) $. The
Rashba vector   $\bnk= \hat {\bf
z} \times \hat \bk $
defines a momentum-dependent spin-quantization axis.

The helicity operator
\begin{eqnarray}
\hat R_{\bk } = \psi \dg_{\bk } ( \bnk \cdot \vec{\sigma })\psi _{\bk }
\end{eqnarray}
commutes with the kinetic part of the Hamiltonian, so that
in the normal state, the quasi-particle basis can be chosen
to be diagonal in the helicity $\beta = \bnk \cdot
\vec{\sigma }$, with corresponding quantum numbers
$\beta = \pm 1$.
The corresponding normal state spectrum is given by $\epsilon_{\bk\pm
}=
\epsilon_{\bk } \pm
\lambda_{\bk} $,  so the spin-orbit term splits the
spin-degeneracy of the Fermi surface (Fig.~\ref{fig:2D SOC Bands} (a)).

\begin{figure}
\begin{center}
\includegraphics[width=\columnwidth]{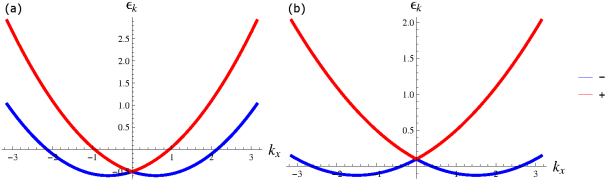}
\end{center}
\caption{2D helical bands respectively with weak spin-orbit coupling, and strong spin-orbit coupling resulting in one of the helical bands pushed above $E_F$.}
\label{fig:Weak Strong SOC}
\end{figure}

The helicity and d-vector define two independent spin quantization
axes. Suppose
first that the Rashba and d-vector rotate with the same (positive) chirality
around the Fermi surface; in this case the angle
$\theta $ between these two axes is constant and  we can write
\begin{equation}
\vec{d} (\bk ) = \cos \theta \bnk + \sin
\theta \hat {\bf k}
\end{equation}
When $\theta  =0$, the two quantization axes align, ${\bf d} (\bk )= \bnk$.
In this case, the pairing and Rashba term commute,
$
[ \hat R_{\bk}, \
\psi\dg _{\bk } ( {\hat  {\bf d} (\bk )}\cdot \vec \sigma) \psi_{\bk }]= 0
$
so helicity becomes a conserved quantum number and
the Bogoliubov quasi-particles acquire a definite helicity.
If we introduce the projection operator onto the helical basis,
\begin{equation}
{\cal P_{\beta } } =
\frac{1}{2}\left(1 +\beta  \bnk \cdot \vec{\sigma
} \right)
, \qquad (\beta = \pm 1)
\end{equation}
then the Hamiltonian can be written
\begin{eqnarray}
\label{eq:He3B helical basis}
{\cal H}_{\bk} & = &
\biggl[ (\epsilon_{\bk } + \lambda_{\bk} )\gamma_{3}
+ \Delta \gamma_{1}\biggl] P_{+}
\cr
&+&\biggl[ (\epsilon_{\bk } - \lambda_{\bk} )\gamma_{3}
- \Delta \gamma_{1}\biggl] P_{+}
\end{eqnarray}
This describes paired Fermi surfaces with ``s-wave'' pair condensates
of opposite sign
and dispersion
\ba
E^{\pm}(\bk) & = & \left[ (\epsilon_{\bk}\pm \lambda_{\bk })^2  + \Delta^2 \right]^{1/2}
\label{eq:Istropic Bogoliubov spectrum}.
\ea

More generally, we can write
\begin{eqnarray}
\label{eq:He3B helical basis2}
{\cal H}_{{\bk }} & = &
[\epsilon_{\bk }+ \lambda_{\bk} \beta_{3} (\bk )] \tau_{3} +
[\Delta_{\parallel \bk } \beta_{3} (\bk ) +\Delta_{\perp \bk }
\beta_{1} (\bk )]\tau_{1} \cr &&
\end{eqnarray}
Here $\beta_{3} (\bk) = \bnk \cdot \vec{\sigma }$ and
$\beta_{1} (\bk ) = \hat \bk \cdot\vec{\sigma}$.
For a positive chirality state
$\Delta_{\parallel \bk }= \Delta \cos \theta$ and
$\Delta_{\perp \bk  }= \Delta \sin\theta$
are the pairing components parallel and perpendicular to the
helicity axis $\bnk$, respectively.
So long as $\cos \theta \neq 0$,
the diagonal component of the gap preserves the $s^{\pm}$ symmetry.
Thus when the the $\bnk$ and $\bdk$ rotate in the same sense,
we obtain a $J = 0$ $d$-wave superfluid ground state.

However, when the  $\bdk$ vectors have a negative chirality,
rotating in the opposite direction to the helicity vector $\bnk$ a
different kind of behavior occurs.
Now
\ba
\label{eq:d vector neg chirality}
\bdk  & = & cos(2 \theta_{\bk} + \phi)\bnk
+ sin(2 \theta_{\bk} + \phi) \hat{\bk}
\ea
where $\theta_{\bk}$ is the azimuthal angle around the Fermi surface, $\theta_{\bk} = \tan^{-1} \tfrac{k_x}{k_y}$,
and $\phi$ is the relative angle between $\bnk $ and $\bdk$ at $\theta_{\bk} = 0$.
The symmetry of the superfluid state is determined by the diagonal,
intra-band component of the pairing in the helical quasi-particle basis,
i.e. $\Delta_{\parallel\bk }$ in Eq.~\ref{eq:He3B helical basis}. From Eq.~\ref{eq:d vector neg chirality},
we see that this is equal to,
\ba
\label{eq:dwave gap}
\Delta^{J = 2}_{\bk} &=  & \Delta cos(2 \theta_{\bk} + \phi)
\ea
Thus when the the $\bnk$ and $\bdk$ counter-rotate,
we obtain a $J = 2$ $d$-wave superfluid ground state.

The full Green's function of the system is given by $\mathcal{G} (z) =
(z - {\cal H}_{\bk })^{-1}$, and the Bogoliubov spectrum is determined by
the poles of $\mathcal{G}$,  which gives,
\ba
E^{\pm}(\bk) & = & \left[ \epsilon_{\bk}^2 + \lambda_{\bk}^2  + \Delta^2 \right. \cr
             &   & \pm 2 \left. \left[ \lambda_{\bk}^2 \epsilon_{\bk}^2 + \lambda_{\bk}^2 \Delta^2 |\hat{n}_{\bk} \times \hat{d}_{\bk}|^2 \right] \right]^{1/2}
\label{eq:Bogoliubov spectrum 1}
\ea
The Bogoliubov spectrum can also be written as,
\ba
E^{\pm}(\bk) =  \left[ A_{\bk} \pm \sqrt{A_{\bk}^2 - \gamma_{\bk}^2} \right]^{1/2}
\label{eq:Bogoliubov spectrum}
\ea
where $A_{\bk}  =  \epsilon_{\bk}^{2} + \lambda_{\bk}^2  + \Delta^2$ and
\ba
\gamma_{\bk}^2  =  (\epsilon_{\bk}^{2} - \lambda_{\bk}^2  + \Delta^2 )^2 + 4 (\lambda_{\bk} \Delta \, \bnk \cdot \bdk)^2
\ea
Since $(E^{+} E^{-})^2 = \gamma_{\bk}^2$, it follows that when
$\bnk \cdot \bdk  \neq 0$,  $E^{+} E^{-}$ is
positive definite, and the gap is finite.  If $\bnk $ and $\bdk $
rotate in the same sense, the gap is finite everywhere and
and maximized when $\bnk$ are $\bdk$ are parallel,
i.e $\bnk \times \bdk = 0$. In a mean-field theory, the system
selects this minimum energy state dynamically,
generating an internal Josephson coupling
that couples the two pairing channels such that the $\bdk$
vector lies parallel to the spin-orbit field $\bnk$.
By contrast, when $\bnk$ and $\bdk $ counter rotate, $\gamma_{\bk}^2 = 0$ along
the nodes of $cos(2 \theta_{\bk} + \phi)$, which is the rotation of a
$d_{xy}$  state through angle $-\phi $. The gap nodes occur at locations where $\gamma_{\bk }=0$, i.e
at the intersection of the nodal lines of $cos(2 \theta_{\bk} + \phi)$
and the surfaces defined by $\epsilon_{\bk}^2 - \lambda^2 + \Delta^2 = 0$.

\section{Topological $s^{\pm}$ \& $d$-wave State with Gapless Edge States}

The topology of $^{3}$He-B is protected by time-reversal symmetry ($\mathcal{T}$)
with an invariant given by Eq.~\ref{eq:winding number}, and there are corresponding
time-reversal protected gapless chiral edge states.
Since the spin-orbit coupling $\bnk \cdot \vec{\sigma}$ is time-reversal invariant,
it will not mix the left and right-moving Majorana fermions.
Furthermore, the system remains fully gapped it is adiabatically evolved from the $^{3}$He-B state into the
$s^{\pm}$ state by switching on the spin-orbit coupling. Hence, we expect the low angular momentum $s^{\pm}$ state to remain topological and exhibit gapless chiral edge states.

For completeness, we calculate the edge states at the domain wall
between two bulk $^{3}$He-R of opposite chirality, satisfying the
boundary conditions $\Delta_2(x = -\infty) = -\Delta_2$ and
$\Delta_2(x = \infty) = + \Delta_2$, using the method described by
\citet{VolovikJTEP2009a}.  This calculation is equivalent to the
calculation of reflection at a boundary along the plane $x=0$ of the
superfluid, since the Rasbha and pairing fields of a fermion
reflecting at normal incidence off the boundary  electron, reverse.
The topological invariant $n$ (Eq.~\ref{eq:winding number}) changes sign
from $+ 1$ to $- 1$ across the domain, when $\Delta(x)$ changes sign.
Similarly, $\lambda(x)$ changes sign so that the system remains in
a $J = 0$ $s^{\pm}$ state on both sides of the domain.
For small $k_x^2 \ll k_F^2$, we can calculate the edge states perturbatively.
Letting $k_x = k_F + i \partial_x$, we obtain the Hamiltonian,
\ba
\label{eq:Edge Hamiltonian}
H & = & H^{(0)} + H^{'} \\
H^{(0)} & = & i v_F \partial_{x} \gamma_3 + \lambda (x) \sigma_2 \gamma_3 + \Delta_2 (x) \sigma_2 \gamma_1 \\
H^{'} & = & \frac{\Delta_1}{k_F} k_y \sigma_1 \gamma_1 + \frac{\lambda}{k_F} k_y \sigma_1 \gamma_3
\ea
where $v_F = \tfrac{k_F}{m}$. There are two zero-energy solutions,
$\psi_{+}$ and $\psi_{-}$ corresponding to $\sigma_{2} = \pm 1$ respectively.
\ba
\label{eq:Zero energy modes}
\psi_{\pm}(x)  &=&  {\mathrm exp} \left[ {-\frac{1}{v_F} \int_0^x dx'
\left(\Delta_2(x') - i \lambda(x') \sigma_2 \right)} \right]
\xi_{\pm},\cr
\xi_{\pm } &=& \bpm 1 \\
\pm i \epm_{\gamma} \!\!\! \bpm 1 \\ \pm i \epm_{\sigma} .
\ea

It is straightforward to show that the zero-energy modes satisfy the following Hamiltonian along the edge, and disperse linearly.
\be
\label{eq:Dirac Hamiltonian}
\begin{bmatrix}
H^{'}_{++} & H^{'}_{+-} \\
H^{'}_{-+} & H^{'}_{--} \\
\end{bmatrix}
=
\begin{bmatrix}
0 & ( v - i \delta )k_y \\
(v  + i \delta )k_y & 0 \\
\end{bmatrix}
\ee
where,
\ba
\label{eq:edge velocity}
v & = & \left( \int_{-\infty}^{\infty} dx \frac{\Delta_1(x)}{k_F} exp[-\frac{2}{v_F} \int_{0}^{x}dx' \Delta_{2}(x')]  \right) \cr
  &   & \times \left( \int_{-\infty}^{\infty} exp[-\frac{2}{v_F} \int_{0}^{x} dx' \Delta_{2}(x') ] \right)^{-1} \\
  & & \cr
\delta & = & \left( \int_{-\infty}^{\infty} dy \frac{\lambda_1(x)}{k_F} exp[-\frac{2}{v_F} \int_{0}^{x}dx' \Delta_{2}(x')]  \right) \cr
       &   & \times \left( \int_{-\infty}^{\infty} exp[-\frac{2}{v_F} \int_{0}^{x} dx' \Delta_{2}(x') ] \right)^{-1}
\ea

Solving the edge Hamiltonian, Eq.~\ref{eq:Edge Hamiltonian}, gives the following two fermionic zero modes,
\ba
\label{eq:Majorana fermions}
H^{'} \psi_{1,2} & = & \pm \; c k_y \psi_{1,2} \cr
\psi_{1,2} & = & \psi_{+} \pm \psi_{1} \cr
c & = & \sqrt{v^2 + \delta^2}
\ea
where $\psi_{1,2}$ are two linearly dispersing Majorana fermions, similar to
the Majorana edge modes found in isotropic $^{3}$He-B, with a renormalization
of the velocity by the spin-orbit coupling.

As explained in Sec.~\ref{sec:He3R}, the $d$-wave state corresponds to counter-rotation
of $\hat{d}_{\bk}$ with respect to $\hat{n}_{\bk}$, and in particular, choosing
$\hat{d}_{\bk} \cdot \vec{\sigma} = \hat{k}_y \sigma_x + \hat{k}_x \sigma_y$ gives a $d_{xy}$ state.
Hence, an identical calculation to that carried out above, with $\Delta_1 \rightarrow - \Delta_1$
shows that the $d$-wave state is also topological with gapless Majorana edges states.
This is in agreement with the results of Schnyder {\it et. al.} \cite{SchnyderPRB2012}.

\section{Effects of Hard-Core/Coulomb Repulsion: Topological Phase Transition Into $d$-wave State}
\label{sec:Coulomb}

The hard-core fermionic repulsion in $^{3}$He requires that the on-site pair amplitude
is zero, $\langle \psi_{\uparrow}(\vec{x}) \psi_{\downarrow}(\vec{x}) \rangle = 0$,
and the $^{3}$He-B phase satisfies this constraint by triplet pairing in the $p$-wave channel.
However, spin-orbit coupling, which allows mixing of spin and angular momentum,
causes scattering of $p_{x/y}$-wave triplet pairs into $s$-wave spin-singlet Cooper pairs,
and this can lead to a finite on-site $s$-wave pair amplitude.

The $s^{\pm}$ state manages to satisfy the hard-core constraint,
even though there is a finite $s$-wave pair susceptibility in each $p$-wave channel,
because of phase cancellation between the bands with opposite helicities.
The phase cancellation mechanism is clear from the Green's function in the helical basis,
which may be calculated from Eq.~\ref{eq:He3B helical basis},
\ba
\label{eq:Greens fn w swave}
{\cal{G} }(z, \bk) & = & \frac{1}{z - \cal{H}_{\bk}} \cr
  & = & \sum_{\pm } \frac{z + (\epsilon_{\bk } \pm \lambda_{\bk}
  )\gamma_{3} \pm \Delta \gamma_{1}}{z^{2} - E^{\pm}(\bk)^2}\biggl(
\frac{1 \pm \bnk \cdot \vec{\sigma} }{2}
\biggr) \cr &&
\ea
The $\pm \Delta \gamma_{1}$ component of the Gork'ov propagator
describes s-wave pairing on the two helicity split Fermi surfaces.
The net $s$-wave amplitude $\langle c\dg_{\uparrow} c\dg_{\downarrow}
\rangle$ is then given by,
\begin{eqnarray}
\label{splussy}
\langle c\dg_{\uparrow} c\dg_{\downarrow}
\rangle &=& \frac{T}{2} \sum_{\bk, n} Tr [{\cal G} (i\omega_{n}, \bk)
\frac{\gamma_{1}}{2}
]\cr
&=& \frac{T}{2} \sum_{ \bk ,i \omega_{n},\beta= \pm  } \beta \frac{\Delta
}{(i\omega_{n})^{2}-E^{\beta} (\bk )^{2}}\cr
&=&
\sum_{\bk , \beta  = \pm }\beta  \tanh \left(\frac{E^{\beta } (\bk
)}{2T} \right)
\frac{\Delta }{4 E^{\beta }_{\bk }}.
\end{eqnarray}
We can interpret the two components in Eq.~\ref{splussy} as the pair
contributions from the two helicity polarized bands, given by
\begin{equation}
\langle c\dg_{\up }c\dg_{\downarrow}\rangle_{\pm } = \pm \sum_{\bk }\tanh \left(\frac{E^{\pm } (\bk
)}{2T} \right)
\frac{\Delta }{4 E^{\pm }_{\bk }}.
\end{equation}
confirming that each band contributes an s-wave pairing amplitude of opposite signs.
In the limit of weak spin-orbit coupling, when $E^{+}(\bk) \approx
E^{-}(\bk)$, there is almost complete phase cancellation between the
two helical bands, and $\langle c\dg_{\uparrow} c\dg_{\downarrow}
\rangle \approx 0$.  However, this mechanism fails when the spin-orbit
coupling becomes comparable to the kinetic energy, such that one of
the bands is shifted away from the Fermi surface; a phase transition
to a $d$-wave state will then occur.

We now include a Hubbard interaction to account for the hard-core repulsion,
and then carry out a Hubbard-Stratonovich decomposition into an $s$-wave term
\ba
\label{eq:Coulomb interaction}
H^{(U)} & = & U \sum_{i} n_{i \uparrow} n_{i \downarrow} \equiv U \sum_{i} (c\dg_{i\up}c\dg_{i\dw }) (c_{i\dw}c_{i\up}) \cr & \longrightarrow & \sum_{i} \Delta_s \, c\dg_{\uparrow i} c\dg_{\downarrow i} + H.c. - \frac{|\Delta_s|^2}{U}
\ea
At the saddle point of the mean-field free energy where $\partial
F/\partial\Delta_{s}=0$, the pair density is given by $\langle
c\dg_{i\up}c\dg_{i\dw } \rangle= \bar \Delta_{s}/U$, and in the large $U$
limit, this becomes the constraint
\be
\hspace{2.5cm} \langle c_{\uparrow} (\vec{x}) c_{\downarrow} (\vec{x}) \rangle = 0  \hspace{1.2cm} (U \rightarrow \infty)
\ee

After including the $s$-wave pairing, the Hamiltonian is now written as,
\begin{eqnarray}
\label{eq:He3B w swave helical basis}
\cal{H}_{\bk} & = & \sum_{\beta= \pm} \biggl[ (\epsilon_{\bk } + \beta \lambda_{\bk} )\gamma_{3}
              + (\beta \Delta + \Delta_s) \gamma_{1} \biggl] P_{\beta}
\end{eqnarray}
and the Bogoliubov spectrum is then given by,
\ba
E^{\beta}(\bk) & = & \left[ (\epsilon_{\bk} + \beta \lambda_{\bk })^2  + (\Delta + \beta \Delta_s)^2 \right]^{1/2}
\label{eq:s-wave spectrum}
\ea

It is now straightforward to calculate the free energy and the $s$-wave amplitude.
\ba
\label{eq:Free energy}
F & = & N_{s}\left[ \frac{\Delta_{1}^{2}}{g_{1}} + \frac{\Delta_{2}^{2}}{g_{2}} \right] \\ \nonumber
  &   & - 2 T \sum_{\bk, \alpha } \ln \left[2 \cosh  \left( \frac{E^{\alpha }_{\bk }}{2T} \right) \right]
\ea

The stationarity condition becomes
\be
\label{eq:swave amplitude}
\frac{\partial F}{\partial \Delta_s} = \langle c_{\uparrow}^{\dg} c_{\downarrow}^{\dg} \rangle =
\langle c^{\dg}_{\uparrow} c^{\dg}_{\downarrow} \rangle_{+} + \langle c^{\dg}_{\uparrow} c^{\dg}_{\downarrow} \rangle_{-} = 0
\ee
where by direct differentiation,  we recover the result of Eq.~\ref{splussy},

\ba
\label{eq:helical swave amplitude}
\langle c\dg_{\up }c\dg_{\downarrow}\rangle_{\pm }
 = \pm \sum_{\bk }\tanh \left(\frac{E^{\pm } (\bk
)}{2T} \right)
\frac{\Delta }{4 E^{\pm }_{\bk }}.
\ea

We now use a simplified momentum-independent spin-orbit coupling $\lambda_{\bk} = \lambda$
to demonstrate the key physics of phase cancellation in 2D.
In this simplified model, the helical bands are split apart by $\lambda$,
and the density of state of both bands remain constant.
The integral in Eq.~\ref{eq:helical swave amplitude} gives the standard BCS result,
\begin{equation}\label{l}
\langle c\dg_{\up }c\dg_{\downarrow}\rangle_{\pm }
= \pm \frac{N(0)\Delta }{2} \ln \frac{\omega_{sf}}{\Delta}
\end{equation}
where $\omega_{sf}$ is
the characteristic upper cutoff of the p-wave pairing attraction (spin-fluctuation) energy scale and $N(0)$ is the density of states.
In this simple case,
 $\langle c^{\dg}_{\uparrow} c^{\dg}_{\downarrow} \rangle_{+}$
and $\langle c^{\dg}_{\uparrow} c^{\dg}_{\downarrow} \rangle_{-}$ exactly cancel.
Thus, in the case of weak to moderate spin-orbit coupling,
when both helical bands still cross $E_F$, there is zero net $s$-wave Cooper pair
amplitude due to phase cancellation of $s^{\pm}$ state on both bands.

However, when the spin-orbit coupling becomes comparable to the kinetic energy
and shifts one of the bands away from $E_F$, there will now be a net $s$-wave pair scattering amplitude.
In the quasi-particle basis, this means that the $s^{\pm}$ state is transformed into an $s^{++}$ state
as there is only one helical band with an $s^{++}$ pairing crossing $E_F$.

\begin{figure}
\begin{center}
\includegraphics[width=\columnwidth]{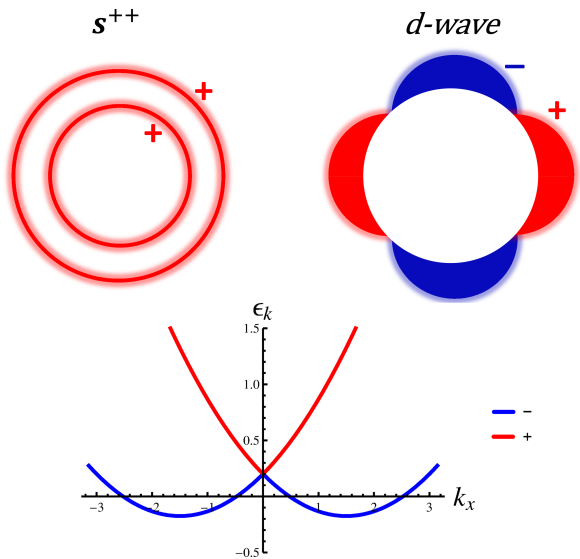}
\end{center}
\caption{Strong spin-orbit coupling scenario: in the absence of Coulomb repulsion,
the $s^{pm}$ state is transformed into an $s^{++}$ state on the remaining helical band.
In the presence of Coulomb repulsion, the on-site $s$-wave pair amplitude is disfavored,
and the system will instead favor a phase transition into a $d$-wave state to minimize the hard-core/Coulomb repulsion.
(Technically, the superfluid pairing on the lower helical band has a phase proportional to $\beta = -$, but we follow convention in labelling it as an $s^{++}$ pairing, which is equivalent up to a gauge transformation.)}
\label{fig:s++ dwave}
\end{figure}

This fully gapped $s^{++}$ state is energetically favored in the
absence of hard-core/Coulomb repulsion.  However, in the presence of a
hard-core/Coulomb repulsion, the finite on-site $s$-wave pair
amplitude is strongly disfavored, and the system will instead undergo
a topological phase transition into a $d$-wave state, as illustrated in Fig.~\ref{Fig:Phase transition}.
The positions of the nodes will be determined by the relative orientation
($\phi$) of the $\hat{d}_{\bk}$-vector with respect to the spin-orbit
$\hat{n}_{\bk}$-vector, and this corresponds to an additional $U(1)$
gauge degree of freedom. For $\phi = 0$, we will get a $d_{x^2-y^2}$
state, while $\phi = \tfrac{\pi}{2}$ will correspond to a $d_{xy}$
state. The $d$-wave state will not be topological, as the gapless
fermionic excitations along the nodes in the $d$-wave superfluid state
will couple to the gapless Majorana edge states in general.

For a realistic momentum-dependent spin-orbit coupling, $\lambda_{\bk} = \lambda |\vec{k}|$,
these results remain qualitatively correct, with corrections due to renormalization of $N(0)$
by the spin-orbit coupling. In this case, the phase cancellation will not be exact, and the phase transition will occur
before the upper helical band is completely lifted above $E_F$.

\section{Discussion}

Using a Rashba coupled model of two dimensional superfluid He-3B,
``$^{3}$He-R", we have demonstrated
that in the presence of a strong Rashba coupling,
a single underlying microscopic pairing mechanism can
give rise to two superfluid/superconducting ground states of different
symmetry : a low ``spin" fully gapped
topological $s^{\pm}$ state, and a high ``spin" gapless $d$-wave
state. This is because the spin and rotational symmetries of a system
are coupled by spin-orbit coupling, i.e. $SU(2)_{S} \otimes
SO(3)_{L} \rightarrow SO(3)_{J}$.

In contrast to previous works \cite{FrigeriSigristPRL2004,SatoSatoshiPRB2009,TanakaSatoPRL2010,SchnyderPRB2012}
on non-centrosymmetric superconductors, where they assumed that
the $\hat{n}_{\bk}$ and $\hat{d}_{\bk}$ vectors are parallel
due to strong spin-orbit coupling, we take into account the additional $U(1)$ rotational
degree of freedom, which gives rise to a low ``spin" to high ``spin" transition.
We show that the strong Coulomb repulsion
breaks the alignment of $\hat{n}_{\bk}$ and $\hat{d}_{\bk}$, and the
mixing of $s$ and $d$-wave spin-singlet, with $p$-wave spin-triplet pairing
naturally arises from the in-phase and counter-phase rotation of
$\hat{n}_{\bk}$ and $\hat{d}_{\bk}$ respectively.
Whereas, the $d$-wave state obtained in previous results
are generated by an $f$-wave triplet pair rotating in-phase
with $\hat{n}_{\bk}$, i.e. a $J = 3 - 1 = 2$ state.

Since the spin-orbit coupling is time-reversal invariant, the
topological nature of the fully gapped $^{3}$He-B state is
protected for weak spin-orbit coupling. In this limit, the ground
state of the system is a fully gapped topological $s^{\pm}$ state, and
we show using an explicit calculation that the gapless Majorana edge
states survive, in agreement with Sato and Fujimoto\cite{SatoSatoshiPRB2009}.

However, on-site Coulomb or hard-core repulsion will drive the system towards a higher angular momentum $d$-wave state when the spin-orbit coupling is sufficiently large to lift one of the helical bands above the Fermi surface. The phase cancellation mechanism that minimizes the on-site $s$-wave pair amplitude for the $s^{\pm}$ state is then no longer effective, and the system will undergo a topological phase transition to a topological $d$-wave state\cite{SchnyderPRB2012}.

Such a topological phase transition may exist
at the boundary between the crust and quantum interior of
neutron stars where the transition from an $s^{\pm}$ to a
$d$-wave superfluid state would be driven by
the rise in short-range repulsion with
increasing density \cite{PinesAlparNature1985}. This would mean
that Majorana fermions already exist in one of the largest superfluid
systems known in nature.

This work also raises the intriguing possibility that the $s^{\pm}$
superconducting state believed to exist in iron-based superconductors
could have a higher angular momentum microscopic pairing mechanism,
which is hidden behind a non-trivial helical quasi-particle
structure. In these systems, the $d_{xz}$ and $d_{yz}$ atomic orbitals
form an iso-spin $(\vec{\alpha})$ representation, which plays a
similar role to spin here, $\vec{\sigma} \leftrightarrow
\vec{\alpha}$. There is a large orbital Rashba coupling in the Fe
systems, $\lambda_{\bk} \sim E_F$, and a microscopic $d$-wave orbital
triplet pairing\cite{OngColemanPRL2013} will give rise to a $J = L +
\alpha = 0$ $s^{\pm}$ state or a $J = 4$ $g$-wave state. This
possibility will be discussed in future work.

We thank Onur Erten for helpful discussions. We also thank Andreas Schnyder and Philip Brydon for pointing out the topological nature of the $d$-wave state, and for very helpful discussions. This work is supported by DOE grant DE-FG02-99ER45790.

\bibliography{He3B}

\end{document}